\documentclass[aps,eqsecnum,prd,twocolumn]{revtex4}
\usepackage{graphics,graphicx}
\usepackage{amsmath}
\usepackage{amssymb,latexsym,mathrsfs}
\usepackage{hyperref}

\def\bea{\begin{eqnarray}}
\def\eea{\end{eqnarray}}
\def\ba{\begin{array}}
\def\ea{\end{array}}

\def\beq{\begin{equation}}
\def\eeq{\end{equation}}

\begin{document}

\title{Weak measurement and Zeno time in a dissipative environment}

\author{Samyadeb Bhattacharya$^{1}$ \footnote{sbh.phys@gmail.com}, Sisir Roy$^{2} $ \footnote{sisir@isical.ac.in}}
\affiliation{$^{1,2}$Physics and Applied Mathematics Unit, Indian Statistical Institute, 203 B.T. Road, Kolkata 700 108, India \\}

\vspace{2cm}
\begin{abstract}

\vspace{1cm}
A generalized expression for weak value of dwell time in dissipative systems has been constructed using the approach of Caldirola and Montaldi. An approximate measure of Zeno time has been found taking an asymmetric double well potential. Atomic tunneling between two surfaces is taken as a practical example. The formalism can be used for any solvable potential with exact or approximate energy eigenvalues.
\vspace{2cm}


\vspace{1cm}
\end{abstract}

\vspace{1cm}

\maketitle

\section{Introduction}
In a recent paper \cite{13} we have estimated the weak value of dwell time for a dissipative spin-half system. This procedure can be used to generalize our result for any solvable potentials with exact or approximate energy eigenvalues. The result can be extended to find a characteristic time scale for the Quantum Zeno effect. This effect (or paradox) is the inhibition of transitions between the quantum states by frequent measurements \cite{14,15,16,17,18,19,20}. Mishra and Sudarshan were the first to call the effect as Quantum Zeno Effect \cite{14}. The short-time behavior of non-decay probability of unstable particle in quantum theory is shown to be not exponential but quadratic \cite{21}. Wilkinson et al. \cite{22} observed this deviation from exponential decay. In 1977, Misra and Sudarshan \cite{14} showed that this behavior when combined with the quantum theory of measurement, based on the assumption of the collapse of the wave function, lead to a very surprising conclusion: frequent observations slows down the decay. An unstable particle would never decay when continuously observed. The very first analysis does not take into account the actual mechanism of the measurement process involved, but it is based on an alternating sequence of unitary evolution and a collapse of the wave function. Apart from presenting a generalized expression of dwell time in dissipative system, the purpose of this paper is to investigate the characteristic features of the short-time nonexponential region of a metastable system. We shall endeavor to give an approximate estimate of the "Zeno" time for dissipative quantum systems in the framework of weak measurement. But in order to do that, it is essential to introduce the concept of tunneling time, especially dwell time. \\
\noindent As of today, we still do not have a good understanding about tunneling time \cite{2,3,4,5,6,7}. The concept of tunneling probability is clear. By contrast, discussions are continuing at the conceptual level for tunneling time. The stationary treatment of tunneling worked very well but the estimated tunneling times (e.g., $10^{-14}$ to $10^{-15}$ sec \cite{8}) were too short to relate them to the experiment. However, the importance of tunneling dynamics has been recognized in many fields even today and the progress of time-resolved spectroscopy has made it possible to observe the phenomena in solids whose time scales are comparable to the tunneling times. We find various tunneling times that are based on different ideas of characterizing the time spent by a particle under the barrier. Hauge and St{\o}veng  \cite{3} mentioned seven different definitions of tunneling time of which the dwell time and the phase time or group delay are reasonably well accepted by the community. Aharonov et.al \cite{9} dealt with the problem of tunneling time from the context of weak measurement.The notion of the weak value of a quantum mechanical observable was originally  introduced by Aharanov et.al \cite{10}-\cite{12}. This quantity is the statistical result of a standard measurement procedure performed upon a pre selected and post selected (PPS) ensemble of quantum systems when the interaction between the measurement apparatus and each system is sufficiently weak. Unlike the standard strong measurement of a quantum mechanical observable which sufficiently disturbs the measurement system, a weak measurement of an observable for a PPS system does not appreciably disturb the quantum system and yields the weak value as the measured value of the observable. Here our motivation is to develop a generalized weak value of dwell time and apply the result to estimate the Zeno time from the context of weak measurement in presence of dissipative environment. We will consider an asymmetric double well potential as a model potential. In Section II we recapitulate the dwell time in dissipative system from the aspect of weak measurement for convenience and deduce the generalized dwell time. In Section III we will elaborate the aspect of Zeno effect and derive the approximate Zeno time from the result of Section II. In Section IV we will consider a specific system for tunneling of an atom between two surfaces with asymmetric double well potential and use the results of previous sections to calculate the Zeno time for this specific case. After that we will conclude with some discussion and possible implications in section V.

\section{Weak value of dwell time in dissipative systems}

It is possible to construct an operator
\beq\label{2.1}
\Theta_{(0,L)} = \Theta(x)-\Theta(x-L)
\eeq
where $\Theta(x)$ and $\Theta(x-L)$ represents Heaviside functions. This operator measures whether the particle is in the barrier region or not.  Such a projection operator is Hermitian and corresponds to a physical observable. It has eigenvalues 1 for the region $0\leq x\leq L$ and 0 otherwise. It's expectation value simply measures the integrated probability density over the region of interest. So the expectation value of $\Theta_{(0,L)}$ divided by the incident flux will represent the dwell time \cite{3,23}. The weak measurement of dwell time means to get the weak value of this mentioned operator. This approach was originally developed by Aharonov et.al \cite{9,10}. Here the measurement interaction between the measuring device and the system is too weak to trigger a collapse of the wave function. Individual measurement has no meaning, since it is too weak to carry any information. But an ensemble average over a sufficiently large number of such individual results are significant. Since the individual measurements of each observable are very imprecise, it is possible to measure non-commuting observables simultaneously, without violating uncertainty principle. So if we are interested in the duration of some process, we can correspond this to a typical weak measurement extended in time, i.e. the interaction between the measuring probe and the system is not impulsive, but has a finite duration. Steinberg has shown \cite{26} that these features make weak measurement theory a very promising background for the study of tunneling time. We directly come to the evaluation of the dwell time without going into the detail, which can be found in \cite{13} and references therein.\\
The time taken by a particle to traverse certain potential barrier is measured by a clock consisting of an auxiliary system which interacts weakly with the particle as long as it stays in a given region. Aharanov et.al \cite{2} considered the interaction Hamiltonian as
\beq\label{2.3}
H_{int}=P_m \Theta_{(0,L)}
\eeq
where $m$ is the degree of freedom.
It is the effective form of the potential, seen by a particle in the $S_z$ state, in the Stern-Gerlach experiment where (0,L) is the region of magnetic field.
The weak value of dwell time is found to be \cite{2}
\beq\label{2.7}
<\tau>^w= \frac{\int_{-\infty}^{\infty}dt\int_0^L \psi_f^{*}(x,t) \psi_i(x,t)dx}{\int_{-\infty}^{\infty} \psi_f^{*}(x,0) \psi_i(x,0)dx}
\eeq
They argued that direct calculation of the dwell time can be made using equation (\ref{2.7}). \\
Now we concentrate on time dependent pre and post selected states with emphasis on decay of excited states considering the time evolution of a quantum mechanical state as
\beq\label{2.8}
|\psi(t)\rangle= U(t-t_0)|\psi(t_0)\rangle
\eeq
where $U(t-t_0)=e^{-iH(t-t_0)}$ is the time evolution operator.\\
\noindent In the light of the time evolution, the weak value of an operator $A$ at a time t, $t_i<t<t_f$, preselected at $t_i$ and postselected at $t_f$, can be defined as \cite{27}
\beq\label{2.9}
A_w= \frac{\langle\psi_f|U^{\dagger}(t-t_f)AU(t-t_i)|\psi_i\rangle}{\langle\psi_f|U^{\dagger}(t-t_f)U(t-t_i)|\psi_i\rangle}
\eeq
The time evolution operator satisfies the unitary and evolution properties
\beq\label{2.10}
UU^{\dagger}=U^{\dagger}U= I
\eeq
and
\beq\label{2.11}
U(t_1-t_2)U(t_2-t_3)=U(t_1-t_3)
\eeq
Let us consider the case of decay of an excited state by considering an initial excited atom coupled to a  bath of $N$ number of other atoms initially in their ground states. Due to the interaction with the bath atoms the concerning system is loosing energy to the bath modes. Choosing the ground state energies of all atoms to coincide and set to be zero, and setting the excited states $E_n$ to satisfy the relation
\beq\label{2.12}
E_n-E_0=n\Delta E,~~~~~0\leq n\leq N
\eeq
it can be shown that the excited states are  equispaced and distributed symmetrically about the excited state of the reference atom, labeled by $n=0$. For the simplicity of the problem, it is assumed that the reference atom is equally coupled to each of the atoms of the bath and so the interaction is described by the real constant Hamiltonian $H$.\\
The Schr\"{o}dinger equation is equivalent to the coupled differential equations
\beq\label{2.13}
\dot{a_0}=-i\sum_n Ha_n e^{-in\Delta Et/\hbar}
\eeq
\beq\label{2.14}
\dot{a_n}=-iHa_0 e^{in\Delta Et/\hbar}
\eeq
where $a_n$ is the amplitude of the excited state. According to Davies \cite{27} equation (\ref{2.13}) and (\ref{2.14}) can be solved exactly by the method of Laplace transformation. The evolution operators $U(t)$ can also be found. If we consider that one atom at a time of the bath is excited, the evolution operator of the relevant sub-space of the full Hilbert space of states will be a $(2N+1)\times(2N+1)$ dimensional matrix, the components of which may be calculated from the equations (\ref{2.13}) and (\ref{2.14}).
\beq\label{2.16}
U_{00}=e^{-\gamma t}
\eeq
in the limit $\Delta E\rightarrow 0$. Here $\gamma$ is the decay constant. Using this limiting solution, from the equations (\ref{2.13}) and (\ref{2.14}) it is found that
\beq\label{2.17}
U_{n0}=iH\left[\frac{e^{-\gamma t+in\Delta Et/\hbar}-1}{\gamma -in\Delta E/\hbar}\right ]
\eeq
which is also in the limit $\Delta E\rightarrow 0$. Using the relation $U^{\dagger}(t)=U(-t)$, we get the time dependent weak value of an operator $A$ as
\beq\label{2.18}
A_w=\frac{\langle \psi_f| U(t_f-t)AU(t-t_i)|\psi_i\rangle}{\langle \psi_f|U(t_f-t_i)|\psi_i\rangle}
\eeq
Without going into further detail, which can be found in our previous work \cite{13}, we state that dwell time can be expressed as
\beq\label{2.19}
\tau_w^D=\int_{t_i}^{t_f} e^{-\gamma(t-t_i)} \left[\frac{1-e^{-2\gamma(t_f-t)}}{1-e^{-2\gamma(t_f-t_i)}}\right]dt
\eeq
Here it is to be noted that the consideration of the weak survival probability as the dwell time conforms with the understanding of tunneling time given by Winful \cite{28,29,30,31,32,33}. Explaining the phenomena of Hartman effect \cite{34} and superluminal barrier transmission, Winful related the tunneling time (specifically ``group delay") with the concept of energy storage and release in the barrier region. He argued that the group delay ($\tau^G$), which is directly related to the dwell time ($\tau^D$) with an additive self-interaction delay ($\tau^I$),
\beq\label{2.20}
\tau^G=\tau^D + \tau^I
\eeq
is actually the lifetime of stored energy (or stored particles) leaking through both ends of the barrier. When the reflectivity is high, the incident pulse spends much of its time dwelling in front of the barrier as it interferes with itself during the tunneling process. This excess dwelling is interpreted as the self-interference delay. Winful successfully disentangled this term from the dwell time \cite{33}. If the surroundings of the barrier are dispersionless, then the self-interference term vanishes, resulting in the equality of the group delay and dwell time. In that case, the dwell time will give a lifetime of energy storage in the barrier region. This interpretation of dwell time is very  important for our further study on Zeno time. Now let us consider the dynamics of dissipation. \\
\noindent The approach we discussed here, to incorporate dissipation in the dynamics of quantum system, was originally developed by Caldirola and Montaldi \cite{35} and Caldirola \cite{36}, introducing a discrete time parameter ($\delta$) that could, in principle, be calculated from the properties of environment such as its temperature and
composition. It is used to construct a retarded Schr\"{o}dinger equation describing the dynamics of the states in the presence of environmentally induced dissipation, which is given by
\beq\label{2.21}
H_i|\psi\rangle=i\hbar\frac{[|\psi(t)\rangle-|\psi(t-\delta)\rangle]}{\delta}
\eeq
Expanding $|\psi(t-\delta)\rangle$ in Taylor series, equation(\ref{2.21}) can be written as
\beq\label{2.22}
H_i|\psi\rangle=i\hbar\frac{[1-e^{-\delta \frac{\partial}{\partial t}}]|\psi(t)\rangle}{\delta}
\eeq
Setting the trial solution as
\beq\label{2.23}
|\psi(t)\rangle=e^{-\alpha t}|\psi(0)\rangle
\eeq
we solve for $\alpha$ to get
\beq\label{2.24}
\alpha=\frac{1}{\delta}\ln(1+i H_i\delta/\hbar)
\eeq
Substituting $\alpha$ in equation(\ref{2.22}) we find that even the ground state decays. To stabilize the ground state, Caldirola and Montaldi \cite{35} rewrite equation (\ref{2.21}) as
\beq\label{2.25}
(H_i-H_0)|\psi\rangle=i\hbar\frac{[|\psi(t)\rangle-|\psi(t-\delta)\rangle]}{\delta}
\eeq
Where $H_0$ represents the ground state. In this case we get
\beq\label{2.26}
\alpha=\frac{1}{\delta}\ln\left(1+i(H_i-H_0)\delta/\hbar\right)
\eeq
Expanding the logarithm upto third order
\beq\label{2.27}
\alpha= \frac{i(H_i-H_0)}{\hbar} + \frac{(H_i-H_0)^2\delta}{\hbar^2} -\frac{i(H_i-H_0)^3\delta^2}{\hbar^3}
\eeq
So the time evolution takes the form
\beq\label{2.28}
\exp\left[-i\left(\frac{(H_i-H_0)}{\hbar}-\frac{(H_i-H_0)^3\delta^2 }{\hbar^3}\right)t-\frac{(H_i-H_0)^2\delta }{\hbar^2}t\right]
\eeq
So from (\ref{2.28}) we can find the decay rate as
\beq\label{2.29}
\gamma=\frac{(H_i-H_0)^2\delta}{\hbar^2}
\eeq
Now if the final hamiltonian is $H_f$, then we can also set
\beq\label{2.30}
\frac{(H_f-H_0)}{\hbar}=\frac{(H_i-H_0)}{\hbar}\left[1-\frac{(H_i-H_0)^2 \delta^2}{\hbar^2}\right]
\eeq
From (\ref{2.30}) we can find
\beq\label{2.31}
\delta=\frac{\hbar}{H_i-H_0}\sqrt{\frac{H_i-H_f}{H_i-H_0}}
\eeq
Putting the value of $\delta$ in (\ref{2.29}) we find that the decay constant takes the form
\beq\label{2.32}
\gamma=\frac{\sqrt{(H_i-H_f)(H_i-H_0)}}{\hbar}
\eeq
Putting this value of $\gamma$ in (\ref{2.19}) and integrating we find the dwell time as
\beq\label{2.33}
\begin{array}{ll}
\tau_w^D=\frac{\hbar}{\sqrt{(H_i-H_f)(H_i-H_0)}} \times\\
~~~~~~~\coth\left(\frac{\tau^M}{2\hbar}\sqrt{(H_i-H_f)(H_i-H_0)}\right)
\end{array}
\eeq
where $\tau^M =(t_f-t_i)$ is the measurement time. Here we arrive at the expression of the generalized dwell time. If we consider the interpretation of Winful, this will give us the lifetime of decaying states in the barrier region. We shall now find an approximate expression for ``Zeno time" in the framework of weak measurement.

\section{Derivation of Zeno time}

Quantum Zeno Effect (QZE) is often associated with the ironic maxim ``a watched pot never boils". The ``watching" here refers to the continuous activity of pulsed measurement. QZE can be theoretically presented in a simple way by considering the short time behavior of the state vector \cite{37}. If $|\psi\rangle$ be the quantum state at $t=0$ and $H$ is the hamiltonian, then the state of the system at time $t$ is $e^{-\frac{iHt}{\hbar}}|\psi\rangle$. The survival probability is
\beq\label{3.1}
P(t)=|\langle \psi|e^{-\frac{iHt}{\hbar}}|\psi\rangle|^2
\eeq
If $t$ is very small, a power series expansion upto $2^{nd}$ order can be given as
\beq\label{3.2}
e^{-\frac{iHt}{\hbar}}=1-\frac{iHt}{\hbar}-\frac{1}{2}\frac{H^2}{\hbar^2} t^2
\eeq
So the survival probability becomes
\beq\label{3.3}
P(t)=|\langle \psi|e^{-\frac{iHt}{\hbar}}|\psi\rangle|^2\approx [1-\frac{(\Delta H)^2}{\hbar^2}t^2]
\eeq
where
\beq\label{3.4}
(\Delta H)^2=\langle \psi|H^2|\psi\rangle-\langle \psi|H|\psi\rangle^2
\eeq
There are many quantum mechanical states with survival probability appearing on ordinary time scales to be decreasing exponential in time. The quadratic time dependence of Eq. (\ref{3.1}) is inconsistent with those states and implies that in such cases Eq. (\ref{3.1}) holds only for very short times. Consider the survival probability $P(t)$, where the interval $[0,T]$ is interrupted by $n$ frequent measurements done on equal interval at times $T/n,2T/n,....T$ . In an ideal scenario, these measurements are nothing but instantaneous projections. The initial state $|\psi\rangle$ of the concerning system is of course an eigenstate of the measurement operator. In that case the survival probability can be given as
\beq\label{3.5}
P(t)\approx \left[1-\frac{(\Delta H)^2}{\hbar^2} (T/n)^2\right]^n
\eeq
which approaches 1 as $n\rightarrow \infty$.\\
\noindent The QZE can be observed as long as the quantum system displays the behavior shown in (\ref{3.5}). From (\ref{3.3}) a time scale can be constructed as
\beq\label{3.6}
\tau^Z=\frac{\hbar}{\Delta H}
\eeq
which is called the ``Zeno time". If the interval between consecutive measurements is smaller than $\tau^Z$ the dynamics is significantly slowed down or even asymptotically halted. QZE has raised and it continue to gather widespread interest \cite{38} mainly because of two reasons: it's foundational implications about the nature of ``quantum measurement" \cite{39}, and it's technological applications since it can be exploited to preserve decoherence free regions \cite{40,41,42}.\\
\noindent Now if $\tau^M$ is the measurement time, then Eq. (\ref{3.5}) can be written as
\beq\label{3.7}
P(t)\approx \left[1-\left(\frac{\tau^M}{\tau^Z}\right)^2\right]^{T/\tau^M}
\eeq
It follows that
\beq\label{3.8}
P(T)\approx \exp\left[-T\tau^M/(\tau^Z)^2\right]
\eeq
The only condition to go from Eq. (\ref{3.7}) to Eq. (\ref{3.8}) is $\tau^M \ll \tau^Z$ ( there is no restriction on $T$ ). So if the lifetime of the decaying state is $\tau^D$, then $P(T)=e^{-T/\tau^L}$, we can define
\beq\label{3.9}
\tau^Z\approx \sqrt{\tau^L \tau^M}
\eeq
Here if we consider the interpretation of dwell time ($\tau^D$) given by Winful \cite{28} as the lifetime of decaying states, then $\tau^L=\tau^D$.
Now recalling Eq. (\ref{2.33})

\beq\label{3.10}
\begin{array}{ll}
\tau_w^D \tanh\left(\frac{\tau^M}{2\hbar}\sqrt{(H_i-H_f)(H_i-H_0)}\right)\\
~~~~~~~= \frac{\hbar}{\sqrt{(H_i-H_f)(H_i-H_0)}}
\end{array}
\eeq
Again we know $\tau^M$ is very small, since the existence of Zeno effect demands frequent successive measurements. If the pre and post selected energy states ($H_i$ and $H_f$) are very closely spaced, then by imposing a condition $\tau^M\ll \frac{2\hbar}{\sqrt{(H_i-H_f)(H_i-H_0)}}$ we get
\beq\label{3.11}
\tau_w^D \frac{\tau^M}{2\hbar}\sqrt{(H_i-H_f)(H_i-H_0)}= \frac{\hbar}{\sqrt{(H_i-H_f)(H_i-H_0)}}
\eeq
which implies
\beq\label{3.12}
\tau_w^D \tau^M=\frac{2\hbar^2}{(H_i-H_f)(H_i-H_0)}
\eeq
So the weak value of Zeno time can be defined as
\beq\label{3.13}
\tau_w^Z=\frac{\sqrt{2}\hbar}{\sqrt{(H_i-H_f)(H_i-H_0)}}
\eeq
Here we arrive at the expression of weak Zeno time considering the dissipative dynamics as expressed by (\ref{2.21}) with any solvable potential for which we can get the energy values.

\section{A particular case of asymmetric double well potential}
In this section we will evaluate the Zeno time considering a particular case of asymmetric double well potential for the tunneling of atom between two surfaces \cite{43}. In contrast to the bulk tunneling of the atom, here the phonons are very strongly coupled and they change the tunneling behavior in a qualitative way. It's an example of ohmic coupling for a two-state system. Some experiments were done in the early 90s \cite{44,45,46,47} to show that scanning tunneling microscopy can be used to demonstrate some fundamental aspects of quantum mechanics, such as the effect of environment in a quantum system. Louis and Sethna \cite{43} showed that how the calculations of macroscopic quantum coherence (MQC) can be tested experimentally via a microscopic system. The potential calculated for the above mentioned system have a double well shape with an asymmetry energy $\epsilon_0$. Taking some calculated parameters from \cite{48}, Louis and Sethna \cite{43} has reported to obtain tunneling elements upto $\Delta/k_B\sim 0.5 K$. Both the tunneling element ($\Delta$) and the asymmetry energy ($\epsilon_0$) can be varied \cite{49}, since by adding an electric field the wells can be biased in either direction. Here we will theoretically investigate a quartic potential \cite{50} to evaluate the corresponding dwell time and Zeno time for the above mentioned process.\\
\noindent Consider a quartic potential of the form
\beq\label{4.1}
V(x)=\frac{1}{2}m\omega_0^2 x^2 \left[\left(\frac{x}{a}\right)^2-A\left(\frac{x}{a}\right) +B\right]
\eeq
where $A$ and $B$ are dimensionless coefficients and $\omega_0$ and $a$ has the dimension of frequency and length respectively. The potential has two minima at
\beq\label{4.2}
x_0=0~~~~~~~\mbox{and}~~~~~~~x_2=\frac{a}{8}\left[3A+\sqrt{9A^2-32B}\right]
\eeq
and these minima are separated by a barrier with maximum at
\beq\label{4.3}
x_1=\frac{a}{8}\left[3A-\sqrt{9A^2-32B}\right]
\eeq
By choosing the parameters $A=14$ and $B=45$, the potential can be shown to have the shape of a asymmetric double potential as depicted in the figure given below.
\begin{figure}[htb]
{\centerline{\includegraphics[width=7cm, height=5cm] {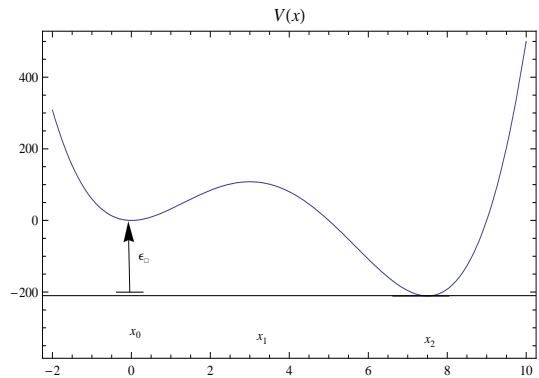}}}
\caption{V(x) vs. x with parameters $A=14$ and $B=45$}
\label{figVr}
\end{figure}

With the dimensionless variable $\xi=x/a$, the potential can be expressed in the form
\beq\label{4.4}
V(\xi)=V_0 \xi^2[\xi^2-A\xi+B]
\eeq
where $V_0=\frac{1}{2}m\omega_0^2a^2$.\\
\noindent The Hamiltonian for this system is
\beq\label{4.5}
H=\frac{p_x^2}{2m}+V(x)
\eeq
where $p_x$ is the momentum conjugate to $x$, hence satisfying the commutation relation
\beq\label{4.6}
[x,p_x]=i\hbar
\eeq
Now the conjugate momentum for the  dimensionless variable $\xi=x/a$ is $p_{\xi}=ap_x$. We also replace $p_{\xi}$ by the dimensionless momentum $p=\frac{p_{\xi}}{m\omega_0 a^2}$. Then the commutation relation takes the form
\beq\label{4.7}
[\xi,p]=\frac{i}{\beta^2}~~~~~~;~~~~\beta^2=\frac{m\omega_0 a^2}{\hbar}
\eeq
In terms of these variables the dimensionless operators, the Hamiltonian takes the form
\beq\label{4.8}
H=V_0\left[p^2+\xi^2(\xi^2-A\xi+B)\right]
\eeq
The Hamiltonian can also be written in the dimensionless form
\beq\label{4.9}
K=\frac{\beta^2}{2}\left[p^2+\xi^2(\xi^2-A\xi+B)\right]
\eeq
where $H=\hbar \omega_0 K$.
To study a gaussian wave packet which is initially located in the left well in such a way that it's maximum is at origin $\xi=0$. The wave packet can pass through the barrier maximum at $\xi=x_1/a$ by tunneling and move to the other well. Expanding the potential given by (\ref{4.4}) around $\xi=0$ and approximating with the harmonic potential (in the dimensionless form) we get
\beq\label{4.10}
V({\xi})=\frac{1}{2}\beta^2B\xi^2
\eeq
Let us now choose the normalized ground state wave function of $V(\xi)$ as
\beq\label{4.11}
\psi(\xi)=\left(\frac{\nu}{\pi}\right)^{1/2}\exp\left[-\frac{1}{2}\nu\xi^2\right]
\eeq
where $\nu=\sqrt{B}\beta^2$. The dimensionless momentum operator is of the form $p=-\frac{i}{\beta^2}\frac{d}{d\xi}$. So the Hamiltonian operator is
\beq\label{4.12}
K=\frac{\beta^2}{2}\left[-\frac{1}{\beta^4}\frac{d^2}{d\xi^2}+\xi^2(\xi^2-A\xi+B)\right]
\eeq
The expectation value of energy in the ground state is found to be
\beq\label{4.13}
K_0=\frac{1}{2}\sqrt{B}+\frac{3}{8\beta^2B}
\eeq
If we displace the center of the wave packet (\ref{4.11}) by a distance $\xi'$, then the expectation value of the energy will become
\beq\label{4.14}
K_0'=K_0 + \xi' \left[\frac{1}{2}\beta^2 \xi' \left(\xi'^2-A\xi'+45\right) \frac{3}{2\sqrt{B}}\left(\xi'-\frac{1}{2}A\right)\right]
\eeq
Therefore the difference in energy in the two minima situated at $0$ and $\xi'=\xi_2$  is
\beq\label{4.15}
h= (K'_0-K_0)\hbar\omega_0=\epsilon \left[1+\frac{\frac{3}{\sqrt{B}}\left(\xi_2-\frac{1}{2}A\right)}{\beta^2 \xi' \left(\xi_2^2-A\xi_2+45\right)}\right]
\eeq
where
\beq\label{4.16}
\epsilon=V(\xi_2)-V(\xi_0)=V_0\xi_2^2(\xi_2^2-A\xi_2+B)
\eeq
For our specific case of double well potential, where $A=14~~ \mbox{and} ~~ B=45$, we get $\xi_2=15/2$ and $\epsilon=\epsilon_0$ which is the asymmetry energy. We also find for this particular case
\beq\label{4.17}
h=\epsilon_0\left(1-\frac{0.0079}{\beta^2}\right)
\eeq
Now it has been shown that the condition $\beta^2>0.0645$ needs to be satisfied for tunneling to occur \cite{51}. Considering this condition, we find that $h\thicksim \epsilon_0$.\\
\noindent Let us now estimate the Zeno time for this particular process (see the figure). Suppose the particle is going from the metastable left well ($K_0\hbar\omega_0$) to the stable right well ($ K_0\hbar\omega_0-|\epsilon_0| $). Here $H_i=K_0\hbar\omega_0,~ H_0=K_0\hbar\omega_0-|\epsilon_0|, ~~\mbox{and}~~ H_f=H_0$. Using Eq. (\ref{3.13}) we find the Zeno time
\beq\label{4.18}
\tau^Z_w=\frac{\sqrt{2}\hbar}{|\epsilon_0|}
\eeq
We calculate the Zeno time for this ``atomic switching" process as mentioned in the beginning of this section, ie the time scale within which the atom will remain on the surface, instead of hopping to the STM tip. The tunneling element can be obtained up to $\Delta/K_B\approx 0.5 K$, ie $0.69\times 10^{-23}$ Joules \cite{43}. So for a slightly biased well with $|\epsilon_0|=\Delta$, we can find the Zeno time from (\ref{4.18}) which is around $21.6$ picoseconds. It is to be noted that Mugnai et. al. \cite{51a} have considered decay from a metastable state in an attempt to suggest that a real time is spent by the system during the process of tunneling. The Zeno time for this process is found to be $30$ picoseconds, which conforms to the time estimated here within the framework of weak measurement.  
\section{Conclusion}
In this work we have estimated the Zeno time for an asymmetric double well potential in a dissipative environment. We have only considered the lowest energy states of the two wells, reducing the system under study effectively to a two-state system. It is necessary to consider the regime of applicability of the discrete two level approximation that we have considered to develop insights about the appearance of Zeno effect in a quantum system. From the result of the preceding sections given by Eq. (\ref{3.13}) and (\ref{4.18}), it is clear that the Zeno time is inversely proportional to the energy values of the particle. So for higher energy states, though the result is very much applicable theoretically, the corresponding Zeno time may be so less that it could be impractical for experimental observations. It is useful to study in the background of weak measurement because for weak measurement, the Zeno time may be much longer than the tunneling time. For very strong measurements, there may not be any Zeno effect as, while such strong measurements still cause small position variance of the wavepacket, the energetic wavepacket moves violently rather than pinning to it's starting point. Here we have taken the case of atomic tunneling between two surfaces as a practical case for evaluating the Zeno time, but the approach can be extended to other cases also. Especially we are interested in the phenomena of macroscopic quantum tunneling systems, where we can extend our method to superconducting current-biased Josephson junction (JJ) \cite{52,53,54,55} or in it's analogs in  ultracold atomic condensates \cite{56}. The potential of the system can also be taken otherwise as per the practical scenario. It is one of the other usefulness of our approach, that as long as we take any exactly or approximately solvable potentials, where we can get the energy eigenvalues, we can use our method to determine the weak value of dwell time or Zeno time without bothering about the shape of the potential. It is worth mentioning that we have been able to extend the application of weak measurement theory in case of atomic tunneling between two surfaces. Our work clearly indicates that weak measurement procedure can be extended to macroscopic quantum mechanical systems, which will be elaborated in subsequent publication.

\end{document}